\documentclass[aps,prb,twocolumn,groupedaddress,showpacs,superscriptaddress,amssymb,amsmath]{revtex4-1}
\usepackage{graphicx}
\usepackage{tabularx}
\usepackage{color}
\usepackage{amsmath}
\usepackage{comment}
\newcommand{\be}{\begin{equation}}
\newcommand{\ee}{\end{equation}}
\newcommand{\bea}{\begin{eqnarray}}
\newcommand{\eea}{\end{eqnarray}}
\usepackage{dcolumn}
\usepackage{hyperref}
\usepackage{bm}
\usepackage{epsf}
\usepackage{subfigure}
\usepackage{epstopdf}%
\usepackage{amsfonts}

\begin{document}

\title{Absence and recovery of cost-precision tradeoff relations in quantum transport} 

\author{Matthew Gerry}
\affiliation{Department of Physics, University of Toronto, 60 Saint George St., Toronto, Ontario M5S 1A7, Canada}

\author{Dvira Segal}
\affiliation{Chemical Physics Theory Group, Department of Chemistry and Centre for Quantum Information and Quantum Control,
University of Toronto, 80 Saint George St., Toronto, Ontario M5S 3H6, Canada}
\affiliation{Department of Physics, University of Toronto, 60 Saint George St., Toronto, Ontario M5S 1A7, Canada}
\email{dvira.segal@utoronto.ca}
\date{\today}

\begin{abstract}
The operation of many classical and quantum systems in nonequilibrium steady state
is constrained by cost-precision (dissipation-fluctuation) tradeoff relations,
delineated by the thermodynamic uncertainty relation (TUR).
However, coherent quantum electronic nanojunctions can escape such a constraint, 
showing finite charge current and nonzero entropic cost with {\it vanishing} current fluctuations.
Here, we analyze the absence, and restoration, of cost-precision tradeoff relations in fermionic nanojunctions under different affinities: voltage and temperature biases.
With analytic work and simulations, we show that both charge and energy currents can display the absence of cost-precision tradeoff if we engineer the transmission probability as a boxcar function---with a {\it perfect} transmission and hard energy cutoffs. 
Specifically for charge current under voltage bias, the standard TUR may be immediately  violated as we depart from equilibrium, and it is exponentially suppressed with increased voltage. 
However, beyond idealized, hard-cutoff energy-filtered transmission functions,  
we show that realistic models with soft cutoffs or imperfect transmission functions follow cost-precision tradeoffs, and eventually recover the standard TUR sufficiently far from equilibrium.
The existence of cost-precision tradeoff relations 
is thus suggested as a generic feature of realistic nonequilibrium quantum transport junctions. 
\end{abstract}

\maketitle 

\section{Introduction}
\label{Sec-intro}

The thermodynamic uncertainty relation (TUR) describes a tradeoff between 
entropy production (cost) and precision (relative fluctuations). 
It allows the detection of nonequilibrium behavior, and further bounds the associated entropy production \cite{Seifert-Rev20,Horowitz20}.
While originally derived in linear response for classical Markovian systems operating at steady state \cite{Barato:2015:UncRel}, it was later proved based on the large-deviation technique 
\cite{Gingrich:2016:TUP,Horowitz:2017:TUR} and an information-theoretic approach \cite{Dechant:2018:TUR}.
The TUR was generalized under different dynamics; a partial list of recent generalizations includes finite-time statistics \cite{Dechant:2018:TUR,Pietzonka:2017:FiniteTUR,Horowitz:2017:TUR,Pigolotti:TURF},
Langevin dynamics \cite{Dechant:2018:TUR,Gingrich:2017,Hasegawa1,TUR-gupta,Hyeon:2017:TUR}, periodic dynamics \cite{Koyuk:2018:PeriodicTUR,Gabri, Esposito20,Potanina},  
and broken time reversal symmetry systems \cite{GarrahanLR-broken,Udo:TURB,Saito,Hyst}. 
TUR relations for integrated currents were derived in Refs. \cite{VanTUR, Landi-PRL}
based on the fundamental fluctuation symmetry for entropy production \cite{fluctRev}. 

Considering quantum systems operating continuously, dissipation-precision tradeoff relations, or more generally, bounds on the precision of a process have been examined in variety of devices. Examples of electrical, thermal and thermoelectric systems are described in, e.g. Refs. \cite{Ptasz,Saito,BijayTUR,Samuelsson,LandiQ,BijayH,Junjie,Hava,Liu:coh,Goold21,Arash,Gernot21,Landi-Boxcar1,Landi-Boxcar2,Gerry21,Gerry22}, demonstrating that the original TUR \cite{Barato:2015:UncRel} can be violated in different operational regimes.
Formally, quantum equations of motion with a Markovian assumption were used to derive bounds on relative fluctuations of observables, particularly considering  the impact of quantum coherences \cite{Carollo19,Hasegawa21a,Hasegawa21b,Saito-OQS}.

The TUR was originally described for multi-affinity systems  in nonequilibrium steady state, and it relied on memoryless (Markovian) dynamics \cite{Barato:2015:UncRel}.
It connected the average current (of charge or energy)  
$\langle j_{\alpha}\rangle$, its variance 
$\langle \langle j_{\alpha}^2\rangle\rangle=\langle j_{\alpha}^2\rangle-\langle j_{\alpha}\rangle^2$, 
and the average entropy production rate $\langle \sigma\rangle $ according to
\bea
\frac{\langle \langle j_{\alpha}^2\rangle \rangle}{  \langle j_{\alpha}\rangle^2} \frac{\langle \sigma\rangle }{ k_B} \geq 2,  \,\,\,\,\,\,\,\,\,\  {\rm TUR}_2
\label{eq:TUR2}
\eea
with $k_B$ the Boltzmann constant. We refer to this bound 
as TUR$_2$, given the lower bound of 2.
Eq. (\ref{eq:TUR2}) provides a lower bound on entropy production in a nonequilibrium steady state process,
that is, a tradeoff between precision and entropy production.

As mentioned above, several studies have exemplified the breakdown of TUR$_2$ in nonequilibrium quantum steady-state devices, with the left hand side of Eq. (\ref{eq:TUR2}) being smaller than 2.
%
A TUR with a two-times looser bound than the original one was derived for quantum systems in a nonequilibrium steady-state
valid up to second order in the thermodynamic affinities \cite{LandiQ}.
This TUR$_1$ bound states 
\bea
\frac{\langle \langle j_{\alpha}^2\rangle \rangle}{  \langle j_{\alpha}\rangle^2} \frac{\langle \sigma\rangle }{ k_B} \geq 1.  \,\,\,\,\,\,\,\,\,\  {\rm TUR}_1
\label{eq:TUR1}
\eea
As was discussed in Refs. \cite{Saito,Gernot21,Landi-Boxcar1}, 
in fermionic systems under voltage bias, the nonequilibrium contribution to the noise 
may reduce the uncertainty such that the cost-precision ratio can
become {\it arbitrarily small}. 
This scenario can be realized by energy filtering a perfect transmission function.
Thus, for fermionic junctions under voltage there is in fact {\it no useful lower bound} on the cost-precision combination, and a trivial relationship (TUR$_0$) holds in the form
\bea
\frac{\langle \langle j_{\alpha}^2\rangle \rangle}{  \langle j_{\alpha}\rangle^2} \frac{\langle \sigma\rangle }{ k_B} \geq 0.  \,\,\,\,\,\,\,  {\rm TUR}_0
\label{eq:TUR0}
\eea
The absence of a cost-precision tradeoff relation was demonstrated in Ref. \cite{Gernot21} 
with simulations for charge current under voltage, by shaping the transmission function through a chain of quantum dots. 
In what follows, we refer to the combination 
$\frac{\langle \langle j_{\alpha}^2\rangle \rangle}{  \langle j_{\alpha}\rangle^2} \frac{\langle \sigma\rangle }{ k_B}$ 
as the TUR ratio (TURR).

In this work, we argue that TUR$_0$ is a theoretical idealized-fragile limit, 
and physical, real quantum transport junctions
must follow nontrivial cost-precision tradeoff relations.
%
Based on analytic expressions and simulations, we show that by using perfect, idealized 
transmission functions one can approach TUR$_0$ in different cases,
either for charge or energy currents, and under different affinities 
(voltage, temperature difference, and both). 
These idealized transmission probabilities 
are in the form of boxcar functions, with {\it sharp} low-energy and high-energy cutoffs and a perfect transmission value. 
However, beyond idealized models, we find that 
realistic transmission functions with either a soft cutoff or an imperfect transmission function observe a cost-precision tradeoff, and eventually recover TUR$_2$ as we increase the thermodynamical forces.
We thus argue that real systems cannot escape cost-precision (entropy production-fluctuations) tradeoff relations,
a generic property of nonequilibrium systems. 


For steady state time-reversible symmetric systems, TUR$_2$ holds in linear response \cite{Barato:2015:UncRel}, also proved in Refs. \cite{BijayTUR, BijayH,Junjie} based on the universal fluctuation symmetry.
This is true irrespective of the underlying dynamics (quantum or classical, Markovian or non-Markovian). Thus, while there are generalizations to the TUR beyond steady state, e.g. by relying on the Lindblad form of the  open quantum system dynamics \cite{Saito-OQS},
a relevant reference point for the investigation of  nonequilibrium steady-state  is TUR$_2$: its potential violation down to TUR$_0$, and its eventual recovery far enough from equilibrium, once physical considerations are taken into account.

The paper is organized as follows. 
In Sec. \ref{Sec-ferm}, we present working formulae for quantum coherent transport.
In Sec. \ref{Sec-voltage}, we approach TUR$_0$ for
charge transport under voltage bias by engineering the  transmission function. We further show that we 
salvage the TUR for realistic models.
The behavior of the TURR for energy currents under voltage bias or a temperature difference 
is examined in Sec. \ref{Sec-ETURR}. Finally, the case with two affinities 
is discussed in Sec. \ref{Sec-two}.
We conclude in Sec. \ref{Sec-summ}.

\begin{figure}[b]
\includegraphics[width=\columnwidth, trim = 20 20 17 20]{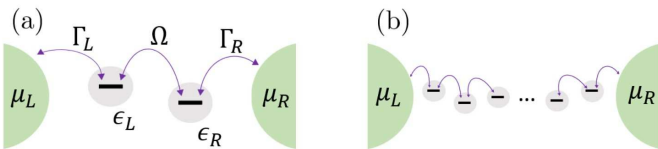}
\caption{
(a) Schematic diagram of a double quantum dot junction. 
(b) A junction consisting of a chain of many quantum dots. It was shown in Ref. \cite{Gernot21} that such chains may be used to engineer transmission functions that approach a boxcar shape.
}
\label{fig:dots}
\end{figure}

\vspace{3mm}
\section{Quantum transport theory}
\label{Sec-ferm}

We consider the problem of noninteracting coherent charge transport with, e.g.,
quantum dots mediating two fermionic leads  
maintained at different chemical potentials and different temperatures. 
The steady-state cumulant generating function (CGF) associated with the 
charge and energy currents is given by \cite{Levitov, fcs-charge1, fcs-charge2}, 
\bea
&&G(\chi_c,\chi_E) = 
\nonumber\\
&&\int_{-\infty}^{\infty} d\epsilon
\ln \Big[ 1 + \tau(\epsilon) \big[ f_L (\epsilon) (1- f_R(\epsilon)) (e^{i (\chi_c+\epsilon\chi_E)}-1) \nonumber \\
&&+ f_R(\epsilon) (1-f_L(\epsilon))(e^{-i (\chi_c+\epsilon\chi_E)}-1) \big]\Big].
\label{eq:CGF}
\eea
%
%
Here, and in what follows, we set the Planck constant, Boltzmann constant, and electron charge to unity, $h=1$, $k_B=1$, $e=1$, respectively.  $\tau(\epsilon)$ is the transmission function for charge carriers at energy $\epsilon$. It
can be calculated from the retarded and advanced Green's function of the system and 
from its self energy matrix \cite{Nitzan,diventra}.
The transmission function is restricted to $0 \leq \tau (\epsilon)\leq 1$. 
$\chi_c$, $\chi_E$ are counting parameters for charge ($c$) and  energy ($E$) transfer processes, respectively, 
in the left terminal.
$f_{\nu}(\epsilon) = 1/\big[\exp(\beta_{\nu} (\epsilon-\mu_{\nu})) + 1\big]$ is the Fermi distribution function for the $\nu=L,R$ metal electrodes. $\beta_{\nu}=1/T_\nu$ is the inverse temperature of the $\nu$ metal with $T_{\nu}$ as the temperature, and $\mu_{\nu}$ is the corresponding chemical potential.

We generate all cumulants from the CGF, particularly the current and its fluctuations. For example, the charge current and its fluctuations are
$\langle j_c \rangle = \frac{\partial G }{\partial (i\chi_c)}|_{\chi=0} $,
$\langle\langle j_c^2 \rangle\rangle = \frac{\partial^2 G}{\partial (i\chi_c)^2}|_{\chi=0} $, respectively.
The resulting currents (positive when flowing left to right) and their fluctuations are given by
\bea
&&\langle j_K\rangle = \int_{-\infty}^{\infty} d\epsilon\;\xi_K\tau(\epsilon) \big[f_L (\epsilon) - f_R(\epsilon)\big],
\nonumber \\
&&\langle \langle j_K^2 \rangle \rangle =
\int_{-\infty}^{\infty} 
d\epsilon\;\xi_K^2 
\Big\{ \tau(\epsilon)  \big[f_L (\epsilon) (1 -f_R( \epsilon))  
\nonumber\\
&&+ f_R(\epsilon)(1-f_L(\epsilon))\big]
- [\tau(\epsilon)]^2  (f_L(\epsilon)-f_R(\epsilon))^2 \Big\},
\label{eq:levitov}
\eea
%
where $\xi_K=1$ ($\epsilon$) for $K=c\:(E)$.

In what follows, we study the TURR analytically and through simulations, with three models for the transmission function:
We use hard-cutoff boxcar transmission functions with cutoff energies $D_{1,2}$ 
and transmission probability $0<\tau_0\leq1$,
\bea
\tau(\epsilon)=
\begin{cases}
\tau_0 & -D_1\leq \epsilon\leq D_2 \\
0 & {\rm else.}
\end{cases}
\label{eq:hard}
\eea
We further exemplify our results with the related function of smooth cutoffs, given by the difference between two sigmoid functions,
%
\bea
\tau(\epsilon)= \tau_0\bigg[
\frac{1}{e^{\gamma(\epsilon-D_2)}+1} -
\frac{1}{e^{\gamma(\epsilon-D_1)}+1}
\bigg].
\label{eq:soft}
\eea
Here, $\gamma^{-1}$ is a broadening parameter (parallel to the role of temperature in the Fermi-Dirac distribution) and $D_1$ and $D_2$ 
are the lower and upper (soft) cutoff energies.
When $\gamma \to \infty$,  Eq. (\ref{eq:soft}) reduces to (\ref{eq:hard}).

We also explore the experimentally-relevant case of electron transport through a serial double quantum dot junction, as illustrated in Fig. \ref{fig:dots}. Such a junction consists of two quantum dots at energies $\epsilon_L$ and $\epsilon_R$, coupled to the left and right leads with coupling strengths $\Gamma_L$ and $\Gamma_R$, respectively, and to each other with tunneling energy $\Omega$. In the symmetric case where $\epsilon_L = \epsilon_R = \epsilon_d$ and $\Gamma_L = \Gamma_R = \Gamma$, the transmission function is given by \cite{BijayTUR}
\be \label{eq:dd_trans}
    \tau(\epsilon) = \frac{\Gamma^2\Omega^2}{|(\epsilon - \epsilon_d + i\Gamma/2)^2 - \Omega^2|^2},
\ee
where choices for the parameters $\Gamma$ and $\Omega$ each contribute to determining its width and the height at which it is peaked.

\begin{figure*} [ht]
\hspace{-6mm}\includegraphics[scale=0.75]{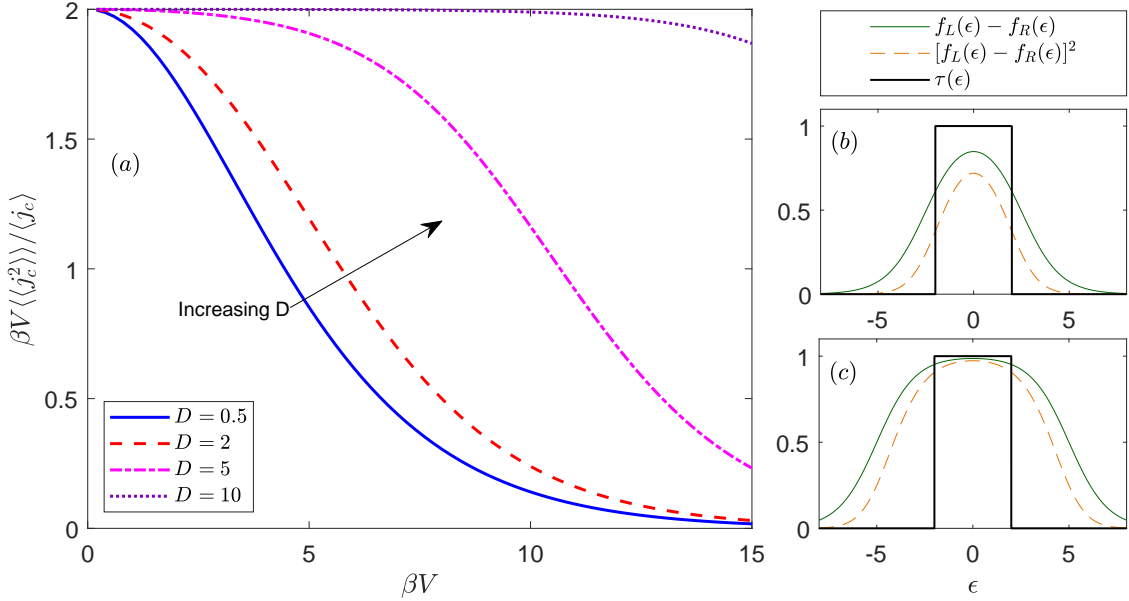}
\caption{
(a) Approaching TUR$_0$ for charge current under voltage 
for a boxcar transmission function with hard cutoffs at $\pm D$.
(b) Difference between the Fermi functions for the left and right metal leads with an applied voltage $V=5$ (full), this difference squared (dashed), and a boxcar transmission probability with $\tau=1$ for $-D\leq\epsilon\leq D$, $D=2$ (thick full). 
(c) The same functions as in (b) at $V = 10$. 
$\beta = 1$ in all calculations here.}
\label{fig1}
\end{figure*}

\vspace{3mm}
\section{Charge transport under voltage}
\label{Sec-voltage}

We begin by studying the behavior of the TURR for a single-affinity 
steady-state charge ($c$) transport under an applied bias voltage $V$.
First, we show in Eq. (\ref{eq:charge_zeroD}) that one can approach TUR$_0$ in an ideal model.
We then argue that physical imperfections restore tradeoff relations,
Eq. (\ref{eq:lambda}),
with the eventual recovery of TUR$_2$, as seen in Eq. (\ref{eq:TUR2highV}). We illustrate this situation with three examples: boxcar-type functions with soft cutoffs or an imperfect transmission coefficient, and a serial double dot junction.

Considering an electronic junction under a voltage bias in steady state, dissipation in the system is given by Joule's heating, 
$\langle \sigma \rangle = \langle j_c\rangle V/T$, with $T$ the temperature
of the electronic system and $\beta= 1/T$.
The TURR 
then simplifies to
\bea
\langle \sigma \rangle \frac{\langle \langle j_c^2\rangle \rangle}{  \langle j_c\rangle^2} =
 \beta V \frac{\langle \langle j_c^2\rangle \rangle}{  \langle j_c\rangle}.
\eea
Based on the relationship 
\bea
&&\left[f_L (\epsilon) (1 -f_R( \epsilon))  + f_R(\epsilon)(1-f_L(\epsilon))\right]
\nonumber\\
&&= 
\coth(\beta V/2)
(f_L (\epsilon)-f_R( \epsilon)),
\label{eq:coth}
\eea
we organize the TURR as
\bea
 \beta V \frac{\langle \langle j_c^2 \rangle \rangle}{\langle j_c \rangle}
& =& \beta V \coth(\beta V/2) 
\nonumber\\
  &-&\beta V \frac{\int_{-\infty}^{\infty} d\epsilon  [\tau(\epsilon)]^2  (f_L(\epsilon)-f_R(\epsilon))^2  }
 {\int_{-\infty}^{\infty} d\epsilon  \tau(\epsilon) (f_L(\epsilon)-f_R(\epsilon))}.
 \label{eq:TUR3}
  \eea
In the absence of the second, nonequilibrium term and 
using $ \coth (\beta V/2)\geq  2/(\beta V)$ we recover TUR$_2$.
It is therefore clear that the second term, which is order of $\tau^2$ and is sometimes referred to as the ``cotunneling" or the ``quantum" contribution, is responsible for potential violations 
of TUR$_2$. 

\subsection{Approaching TUR$_0$ in an idealized model}
\label{Sec-charge-ideal}

As discussed in Refs. \cite{Gernot21,Landi-Boxcar1}, TUR$_0$ can be approached for charge current under voltage by energy filtering the transmission function. We
now first provide an intuitive picture for this result, then a rigorous analysis.

The ratio of integrals  $ \frac{\int_{-\infty}^{\infty} d\epsilon  [\tau(\epsilon)]^2  (f_L(\epsilon)-f_R(\epsilon))^2  }
 {\int_{-\infty}^{\infty} d\epsilon  \tau(\epsilon) (f_L(\epsilon)-f_R(\epsilon))}$,
 which builds up the charge current noise,
can approach the value 1 when the band is structured such that 
 the transmission probability is nonzero within a finite range only
$[-D,D]$, and zero elsewhere as in Eq. (\ref{eq:hard}). Furthermore, we require that the voltage is high, such that
$\beta V\gg1$  and $V>D$. 
As a result,
 $f_L-f_R$ is approximately 1 when the transmission is nonzero. 
Furthermore, if we  require that the nonzero transmission approaches 1 
we get that $(f_L-f_R)^2\approx (f_L-f_R)$ and $[\tau(\epsilon)]^2\approx \tau(\epsilon)$,
 allowing in this limit a full cancellation of the second term in the right hand side in Eq. (\ref{eq:TUR3}) by the first term.
 
We illustrate this argument, on the approach to TUR$_0$ for charge current at high voltage, in Fig. \ref{fig1}.
It is clear from  Fig. \ref{fig1}(a) that at high enough voltage, $V\gg D$, one approaches TUR$_0$.
To rationalize this behavior, we focus for example on the $D=2$ case.  We note from Figs. \ref{fig1}(b)-(c) 
that only for large enough voltage, $(f_L-f_R)^2 \approx (f_L-f_R)$ within the transmission window.
Together with $\tau= \tau^2=1$, a substantial cancellation of fluctuations takes place, as described above. 
Next, we discuss this behavior in a quantitative manner.


\begin{figure*}[t!]
\begin{center}
\includegraphics[scale=0.65,trim=50 10 50 5]{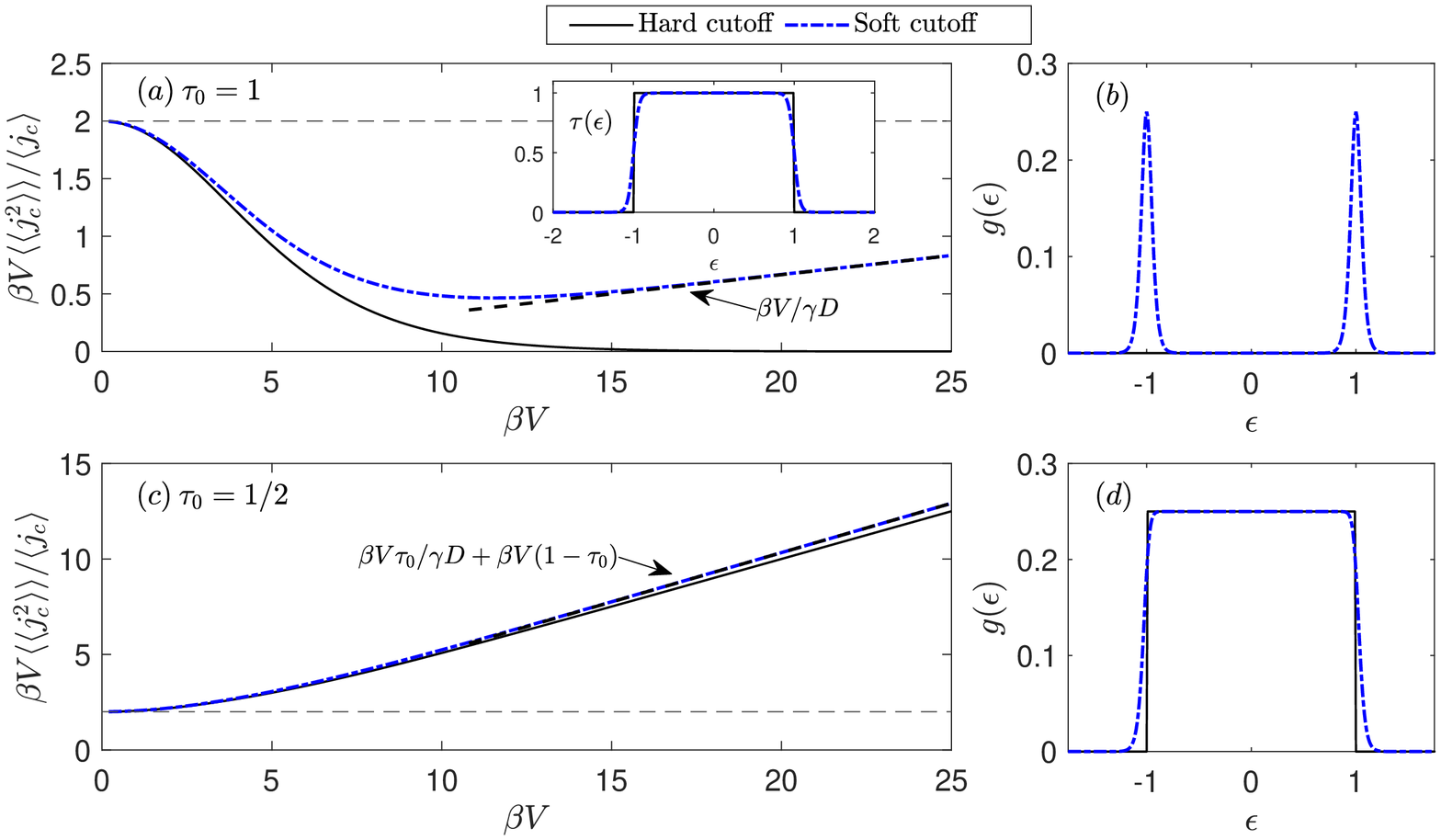}
\caption{
Recovery of a cost-precision tradeoff with nonideal transmission functions.
(a) The TURR for charge transport as a function of bias voltage for both a hard cutoff (full) and a soft cutoff (dashed-dotted)
with a decay parameter $\gamma=30$.
Other parameters are $\tau_0 = 1$, $\beta = 1$, and $D=1$. Inset: the associated hard and soft cutoff transmission functions. (b) The energy-resolved fluctuations at high voltage ($V=25$) with a soft cutoff. This function effectively vanishes everywhere with a hard cutoff when $\tau_0=1$. (c) The TURR with all parameters the same as in (a), except $\tau_0=1/2$. Violations of TUR$_2$ are not observed at low voltage, and the high voltage behavior is predominantly given by the term proportional to $1-\tau_0$ since $\gamma D\gg1$. (d) The associated energy-resolved fluctuations at $\tau_0=1/2$, now similar for a hard and soft cutoff, with diminished significance of the contribution from the regions where $\epsilon\approx \pm D$.
}
\label{fig:soft_cutoff}
\end{center}
\end{figure*}

First, we examine the case of a constant transmission function (not necessarily perfect) extending between $\pm \infty$,
$\tau(\epsilon)=\tau$, $0<\tau\leq1$. In this case, 
\bea
\tau^2\int_{-\infty}^{\infty} d\epsilon    \left[f_L(\epsilon)-f_R(\epsilon)\right]^2 & =
& \tau^2 [-2/\beta +V\coth(\beta V /2)],
\nonumber\\
\tau \int_{-\infty}^{\infty} d\epsilon  \left[f_L(\epsilon)-f_R(\epsilon)\right]&=&  \tau V,
\label{eq:eq10}
\eea
and the TURR for charge transport under voltage bias is
\bea
 \beta V \frac{\langle \langle j_c^2 \rangle \rangle}{\langle j_c \rangle}&=&
 \beta V \coth (\beta V/2) 
  \nonumber\\ 
  &-&  \beta V\tau[-2/(\beta V) +\coth(\beta V /2)]
  \nonumber\\
  &=& \beta V \coth (\beta V/2)  (1-\tau) + 2\tau\geq2.
  \label{eq:eq11}
\eea
TUR$_2$ is therefore valid in the constant transmission limit at any voltage \cite{Hava}.

Next, we consider the opposite, hard cutoff case 
with $\tau(\epsilon)=\tau_0$ for $-D\leq\epsilon\leq D$ but zero elsewhere.
Assuming that $\tau_0=1$, we find
\begin{subequations}
\begin{align}
&&\langle j_c \rangle=
2T \ln\left[ \frac{\cosh\left(\frac{2D+V}{4T}\right)}
{\cosh   \left(\frac{-2D+V}{4T}\right)    } \right], 
\label{eq:jcV} 
\\
&&\langle\langle j_c^2 \rangle\rangle=
%
2T\left[\frac{\sinh(D/T)}{\cosh(D/T) + \cosh(V/2T)}\right].
\label{eq:varcV}
\end{align}
\end{subequations}
%
%
We now simplify the TURR in various limits.

\begin{figure*}[t]
\begin{center}
\includegraphics[scale=0.65,trim=70 10 70 5]{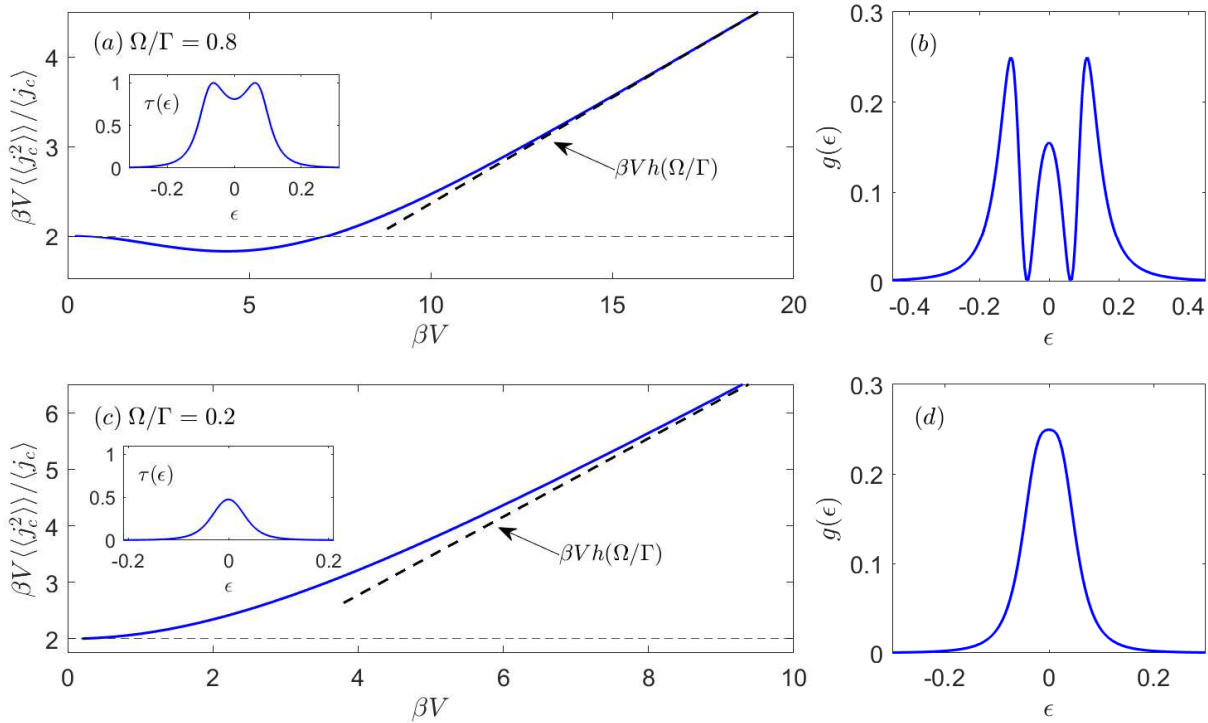}
\caption{
Cost-precision tradeoffs with nonideal transmission functions, focusing on
the recovery of TUR$_2$ far from equilibrium for electronic transport through a serial double quantum dot junction. (a) The TURR for charge transport as a function of bias voltage with $\beta=1$, $\Gamma = 0.1$, and $\Omega=0.08$. (b) The energy-resolved fluctuations at high voltage ($V$=20) with $\Omega=0.08$. (c) The TURR with all parameters the same as in (a) except $\Omega=0.02$; the diagonal asymptote has a correspondingly steeper slope than in panel (a). (d) The energy-resolved fluctuations at $V=20$ with $\Omega = 0.02$. Insets: the transmission functions $\tau(\epsilon)$ associated with the curves shown.
}
\label{fig:TUR_dd}
\end{center}
\end{figure*}

{\it High voltage.} Considering Eqs. (\ref{eq:jcV})-(\ref{eq:varcV}) in the limit of $V\gg D$, the charge current approaches a constant value, $2D$,
while the noise is exponentially suppressed with voltage. 
As a result, the TURR follows
\bea
\label{eq:charge_zeroD}
 \beta V \frac{\langle \langle j_c^2 \rangle \rangle}{\langle j_c \rangle}
&\xrightarrow{V/D\gg1}&\frac{\beta V}{1 + \cosh(\beta V/2)},
\eea
which, for sufficiently large $\beta V$, becomes
\be\label{eq:charge_zeroD}
 \beta V \frac{\langle \langle j_c^2 \rangle \rangle}{\langle j_c \rangle}
\xrightarrow{\beta V\gg1}2\beta Ve^{-\beta V/2}.
\ee
Therefore, when the bandwidth and temperature are small relative to the applied voltage, and despite the fact that the charge current itself does not vanish, 
the noise is exponentially suppressed with voltage, and we approach TUR$_0$. 

{\it Low voltage.} In the low-voltage limit, we perform a series expansion of the TURR in powers of $\beta V$ using the charge current and its fluctuations from 
Eqs. (\ref{eq:jcV})-(\ref{eq:varcV}), respectively. 
In the case that $\tau_0 = 1$ we get
%
\bea
\beta V\frac{\langle \langle j_c^2 \rangle \rangle}{\langle j_c \rangle} 
= 2 - \frac{1}{6\left[1 + \cosh(\beta D)\right]}(\beta V)^2 + O((\beta V)^4).
\nonumber\\
\eea
%
Furthermore, in the limit of large $D$, we get, to the quadratic order in $\beta V$,
\bea
 \beta V \frac{\langle \langle j_c^2 \rangle \rangle}{\langle j_c \rangle} \to 2
- \frac{(\beta V)^2}{12}e^{-\beta D}.
\label{eq:TURRjcEq}
\eea
Thus, TUR$_2$ is {\it immediately} violated when using the boxcar model for transmission functions.
However, this breakdown exponentially 
diminishes as we increase the bandwidth, and eventually we recover Eq. (\ref{eq:eq11}) (here with $\tau=1$).

We illustrate the approach to TUR$_0$ for charge current under a high bias 
voltage in Fig. \ref{fig1}(a) 
using a boxcar function for the transmission probability. 
The figure further illustrates that 
TUR$_2$ is continuously violated with voltage as the relative importance of the $\tau^2$-term grows with increasing voltage.

\subsection{Recovering the TUR: Soft cutoffs and imperfect transmission}\label{sec-soft}

In Sec. \ref{Sec-charge-ideal}, we showed the approach to TUR$_0$ at high voltage
once two conditions were met: (i) the transmission function was perfect ($\tau_0=1$) and (ii) it was trimmed with a hard cutoff. 
We now examine the behavior of the TURR under similar circumstances, but in the realistic scenarios of 
a transmission function either being imperfect or exhibiting soft energy cutoffs. 
We show that in these cases, a nontrivial tradeoff relation holds, 
and TUR$_2$ is always recovered at high enough voltage.

The general argument for the restoration of tradeoff relations for physical systems goes back
to Eq. (\ref{eq:TUR3}). We note that in our convention for biases, $[f_L(\epsilon)-f_R(\epsilon)] \geq [f_L(\epsilon)-f_R(\epsilon)]^2$ and that  $\tau(\epsilon)\geq [\tau(\epsilon)]^2$.
Thus, an imperfection that reduces the transmission function, even if only in a small interval, would result in,
\bea
 \beta V \frac{\langle \langle j_c^2 \rangle \rangle}{\langle j_c \rangle}
 = \beta V [\coth(\beta V/2) -\lambda(V)] \geq 0,
 \label{eq:lambda}
  \eea
with $\lambda(V)<1$ collecting the ratio of integrals in Eq. (\ref{eq:TUR3}).
Eq. (\ref{eq:lambda}) points to the generic existence of a cost-precision tradeoff relation in {\it physical systems}. Note that since $\lambda(V)$ may show a nontrivial voltage dependence, the overall trend in the above relation is not necessarily linear in voltage. 
We now discuss this fundamental result with concrete models, and further show that TUR$_2$ is recovered far from equilibrium, that is, at high voltage.

{\it High voltage.}
Provided that $\tau(\epsilon)$ is integrable, the limit that $V\gg D$ corresponds broadly to a unidirectional transport with
$f_L(\epsilon) \approx 1$ and $f_R(\epsilon) \approx 0$ for any value of 
$\epsilon$ at which $\tau(\epsilon)$ is substantially greater than $0$. 
In this case, expressions for $\langle j_c\rangle$ and $\langle\langle j_c^2\rangle\rangle$, 
given by Eq.~(\ref{eq:levitov}), simplify greatly. 
As in the case of a hard cutoff, the charge current saturates in the large $V$ limit. 
The fluctuations converge to a constant value given by
\be
    \langle\langle j_c^2\rangle\rangle\xrightarrow{V\gg D}\int_{-\infty}^\infty d\epsilon\;\tau(\epsilon)(1-\tau(\epsilon)),
\ee
where the high voltage assumption permits the bounds of integration to be taken to $\pm\infty$. Clearly, for a boxcar function with a perfect transmission up to a hard cutoff, or any combination of such functions \cite{Landi-Boxcar1}, this describes vanishing fluctuations and consequently a vanishing TURR as described by Eq.~(\ref{eq:charge_zeroD}). However, so long as there exists {\it any} range of $\epsilon$ for which $0<\tau(\epsilon)<1$, the fluctuations do not vanish, and the TURR exhibits proportionality with $V$, as
\be 
    \beta V\frac{\langle\langle j_c^2\rangle\rangle}{\langle j_c\rangle}\xrightarrow{V\gg D}\beta V\frac{\int_{-\infty}^\infty d\epsilon\;\tau(\epsilon)(1-\tau(\epsilon))}{\int_{-\infty}^\infty d\epsilon\;\tau(\epsilon)}.
    \label{eq:TUR2highV}
\ee
A central outcome is thus that imperfect 
transmissions would eventually lead, at high enough voltage, to the validity of the 
TUR$_2$ cost-precision relation. We now exemplify this result with three cases.

{\it Case I: Soft cutoffs.}
Considering specifically the transmission function given by Eq.~(\ref{eq:soft}), with $\tau_0=1$, 
soft cutoffs at $\pm D$, and a finite broadening parameter, $\gamma^{-1}$, 
Eq. (\ref{eq:TUR2highV}) reduces to
\bea\label{eq:TUR_soft}
    \beta V\frac{\langle\langle j_c^2\rangle\rangle}{\langle j_c\rangle} &\xrightarrow{V\gg D}&\beta V\bigg[\frac{1}{\gamma D} + 1 - \coth(\gamma D)\bigg]
    \nonumber\\
    &\xrightarrow{\gamma D\gg1}& \frac{\beta V}{\gamma D}.
\eea
Thus, we clearly observe that for a physical transmission function with a non-infinite $\gamma$, that is, having a certain
scale for the softness of the cutoff, TUR$_2$ is recovered at high voltage.

The linear dependence on $V$ as predicted by Eq. (\ref{eq:TUR_soft})
is shown in Fig.~\ref{fig:soft_cutoff}(a). For a nearly rectangular transmission function, when $\gamma D\gg 1$, 
the TURR takes on the form of a simple ratio, with the form of the denominator arising from the fact that the current scales 
with $D$, while the fluctuations scale with $\gamma^{-1}$. 
The existence of a value of $V$ beyond which the TURR has a value greater than 2 is guaranteed, 
ensuring that TUR$_2$ is recovered sufficiently far from equilibrium.

{\it Case II: Imperfect transmission.}
For the same transmission function, Eq. (\ref{eq:soft}), but in the case that $\tau_0$ is strictly less than 1, qualitatively different behavior is observed. 
In this case, Eq.~(\ref{eq:TUR_soft}) generalizes to
\be\label{eq:TUR_soft2}
 \beta V\frac{\langle\langle j_c^2\rangle\rangle}{\langle j_c\rangle} \xrightarrow{V\gg D} \beta V\bigg[\frac{\tau_0}{\gamma D} + 1 - \tau_0\coth(\gamma D)\bigg].
\ee

%
Again, the TURR grows linearly with $V$ at high voltage, guaranteeing TUR$_2$ is satisfied sufficiently far from equilibrium, 
as shown in Fig. \ref{fig:soft_cutoff}(c). 
However, the extra factor of $\tau_0<1$ prevents the cancellation of the last two terms in Eq.~(\ref{eq:TUR_soft2}). 
Instead, we see in Fig. \ref{fig:soft_cutoff}(d) that fluctuations have a contribution dependent on the bandwidth, $D$, as well as dependent on the decay factor,  $\gamma$.
For sufficiently small $\tau_0$ ($\tau_0\lesssim 1/2$) and for large $\gamma D$, 
the $\gamma$-dependent contribution to the noise (at the edge of the band) may be neglected, and Eq.~(\ref{eq:TUR_soft2}) simplifies to match the TURR for a transmission function with hard energy cutoffs,
\be 
    \beta V \frac{\langle\langle j_c^2\rangle\rangle}{\langle j_c\rangle}\xrightarrow{V\gg D, \gamma D\gg1} \beta V(1 - \tau_0).
\ee
%
%
TUR$_0$ can be obtained for a perfect boxcar transmission function, when $\tau_0=1$ {\it and} $\gamma\to \infty$. 
However, any deviation from these settings, e.g. having a finite broadening parameter $\gamma^{-1}$ eventually leads to the recovery of
TUR$_2$ sufficiently far from equilibrium.

{\it Case III: quantum dot setup.}
Similar behaviour is observed at high voltage for transport through a double quantum dot, whose transmission function is given in Eq.~(\ref{eq:dd_trans}). In this limit, the noise and current both saturate, and the TURR takes on the form,
\be 
    \beta V \frac{\langle\langle j_c^2\rangle\rangle}{\langle j_c\rangle} \xrightarrow{V\gg \Gamma, \Omega}\beta V h\left(\frac{\Omega}{\Gamma}\right),
\ee
with
\be 
    h(x) = \frac{1 - 2x^2 + 8x^4}{1 + 8x^2 + 16x^4}.
\ee
TUR$_2$ is thus guaranteed satisfied at sufficiently high voltage ($h(x)>0$ for all $x$). The steepest increase of the TURR with voltage occurs as $\Omega/\Gamma \rightarrow 0$. In this limit the transmission is suppressed at its peak, as shown in the inset of Fig. \ref{fig:TUR_dd}(c). For larger $\Omega/\Gamma$, violations of TUR$_2$ may be observed before it is recovered at higher voltage. It is in this regime that, in analogy to the case of the nearly rectangular transmission function discussed above, the transmission function attains values closer to 1, and features of the energy-resolved fluctuations, $g(\epsilon)$, away from $\epsilon=0$ contribute more substantially to the behavior of the TURR (see Fig. \ref{fig:TUR_dd}(a) and (b)). 

The ultimate recovery of TUR$_2$ far from equilibrium was observed in simulations reported in Ref. \cite{Gernot21}. There, a boxcar function was approached using a carefully engineered chain of quantum dots. However, given small deviations from the perfect boxcar shape, TUR$_0$ could not be absolutely achieved, and TUR$_2$ eventually soared with a linear growth with voltage. Our analytical work here expounds these observations. 


%
{\it Low voltage.}
We now analyze the low-voltage case, $V<D$. 
The behavior of the TURR in the case of a soft energy cutoff is similar to that of a hard energy cutoff, as shown in Fig.~\ref{fig:soft_cutoff}. 
This is attributed to the fact that the integrals of Eq.~(\ref{eq:levitov}) take the bulk of their contributions  from the region where $-V/2<\epsilon<V/2$. 
In this region, the transmission function is approximated well as a constant at $\tau_0$, and the nature of the cutoff is insignificant. Thus, violations of TUR$_2$ are observed at low voltage. 
These violations are the largest when $\tau_0=1$, and they decrease in magnitude for $\tau_0<1$ 
due to the reduced cancellation of the $\tau(\epsilon)$- and $\tau^2(\epsilon)$-dependent contributions to the fluctuations, as discussed in Sec. \ref{Sec-charge-ideal}. Violations of TUR$_2$ are also observed at low voltage for the serial double quantum dot, as discussed in Ref. \cite{BijayTUR}.

\begin{figure*}[htpb]
\begin{center}
\includegraphics[scale=0.6,trim=50 20 50 5]{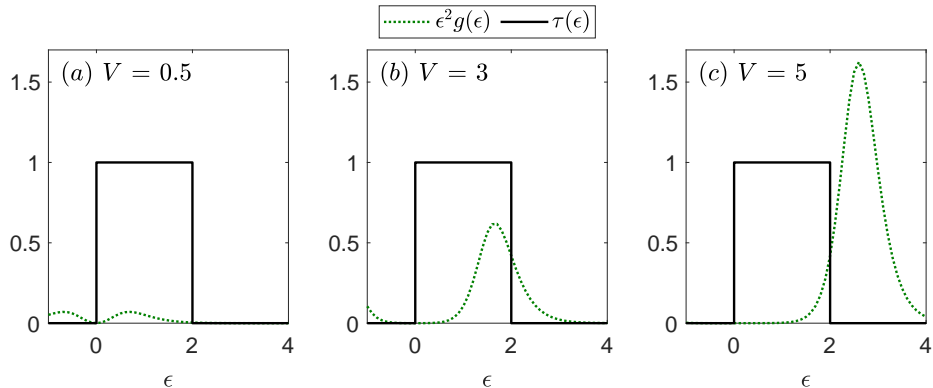}
\caption{Integrand of the expression for the fluctuations in the energy current under various values of bias voltage, compared with the``half-rectangular" transmission function. $\beta = 4$, $\tau_0 = 1$ and $D = 2$.
As we increase voltage, $\epsilon^2 g(\epsilon)$ first arises, then declines within the transmission window.}
\label{noise_integrand_V}
\end{center}
\end{figure*}

\begin{figure}[h]
\includegraphics[width=1.08\columnwidth]{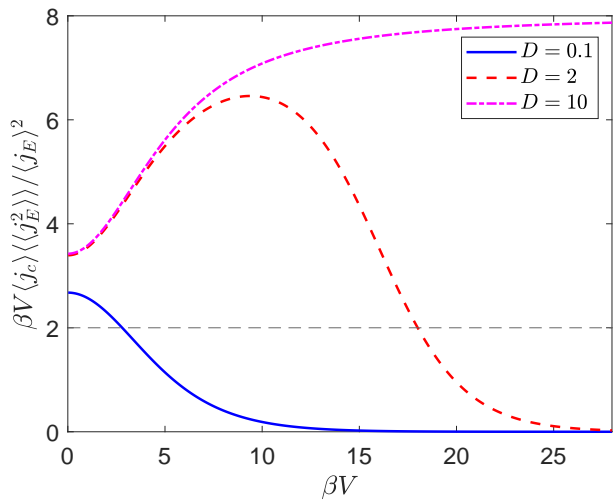}
\caption{TUR ratio for energy transport due to a bias voltage as a function of $\beta V$, $\beta=4$. At low $V$, TUR$_2$ is satisfied, but for sufficiently large $V$ the TURR approaches zero. We present results for narrow (full),
intermediate (dashed) and wide (dashed-dotted) bands, comparing $D$ to voltage.
For $D=2$ the non monotonous trend, from $\beta V$=2, to 12, and 20 can be explained based on Fig. \ref{noise_integrand_V} where energy fluctuations are presented.
}
\label{fig_voltage_etur}
\end{figure}

\section{Energy transport} 
\label{Sec-ETURR}

We continue and study here the TUR for energy transport under voltage bias or a temperature difference.
For the former, we demonstrate that one can approach TUR$_0$ by designing an energy-filtered hard-cutoff, perfect transmission function, 
similarly to the charge transport case \cite{Landi-Boxcar1}.
In contrast, when the energy current is driven by a temperature difference, we show that it is impossible to approach TUR$_0$, though one can violate TUR$_2$.
However, similarly to the charge-current case, deviations from the ideal setting, e.g. by adding a soft cutoff to the boxcar function, lead to the restoration of a nontrivial tradeoff relation,
and to the eventual recovery of TUR$_2$ far enough from equilibrium.

\subsection{Energy transport under voltage bias}  

Identifying the entropy production rate in steady state by the Joule's heating,
$\langle \sigma \rangle = \langle j_c\rangle V/T$,  
we get a compound expression for the  TURR, which involves both charge and energy currents,
\bea
\langle \sigma \rangle \frac{\langle \langle j_E^2\rangle \rangle}{  \langle j_E\rangle^2} 
= \beta V \langle j_c\rangle  \frac{\langle \langle j_E^2\rangle \rangle}{  \langle j_E\rangle^2}.
\label{eq:TURE}
\eea
Returning to (\ref{eq:levitov}), using (\ref{eq:coth}), we get 
\begin{widetext}
\bea
\beta V \langle j_c\rangle  \frac{\langle \langle j_E^2\rangle \rangle}{  \langle j_E\rangle^2} &=& 
 \beta V 
 \int_{-\infty}^{\infty} d\epsilon \tau(\epsilon) \left[f_L (\epsilon) - f_R(\epsilon)\right]
 \nonumber\\
 &\times&
\frac{ 
\left\{ \coth (\beta V/2)
\int_{-\infty}^{\infty} 
d\epsilon\ \epsilon^2 \tau(\epsilon) 
 (f_L(\epsilon)-f_R(\epsilon)) - 
\int_{-\infty}^{\infty} 
d\epsilon\ \epsilon^2 [\tau(\epsilon) ]^2
 (f_L(\epsilon)-f_R(\epsilon))^2 \right\}}
 {\left[\int_{-\infty}^{\infty} 
d\epsilon\ \epsilon \tau(\epsilon) 
 (f_L(\epsilon)-f_R(\epsilon))\right]^2}.
 \label{eq:28}
\eea
\end{widetext}
In ideal settings, $\tau(\epsilon)=[\tau(\epsilon)]^2$. Therefore, at high voltage $\langle \langle j_E^2\rangle\rangle \to 0$ and there is no tradeoff relation---as long as the energy current in the denominator is finite.
However, taking into account small imperfections in the transmission function, whether in the form of a soft cutoff or making it smaller than 1 in a certain region, it is clear than the energy current fluctuations (numerator) will be greater than zero (since $\coth x>1$ for $x>0$, and $0<(f_L(\epsilon)-f_R(\epsilon))\leq 1$), leading to the restoration of a nontrivial tradeoff relation {\it at any nonzero voltage}. 

Next, we examine the combination (\ref{eq:28}) in a quantitative manner, and identify the growth of the TURR with voltage at high voltage.
Note that, if we choose, as before, a transmission function that is 
symmetric around zero (the equilibrium Fermi energy), 
the mean energy current would vanish due to odd symmetry of the integrand in the integral expression, Eq. (\ref{eq:levitov}). 
In contrast, the additional factor of $\epsilon$ in the integrand for the energy current variance results in an 
{\it even} symmetry, thus the variance is finite. 
As a result, a symmetric choice of a boxcar function for $\tau(\epsilon)$, and at any value of $V$, 
leads to a diverging TURR, so TUR$_2$ is satisfied.

We return to a transmission function with a hard energy cutoff. To eliminate this symmetry, 
we consider instead a ``half-rectangular" function with $\tau(\epsilon) = \tau_0$ for $0\leq\epsilon\leq D$ and 
zero otherwise, see Fig. \ref{noise_integrand_V}.
The energy current and its fluctuations are given by Eq.~(\ref{eq:levitov}) with $\xi_K=\epsilon$. 
For $\tau_0=1$, the energy current noise is bounded from above,
%
\bea
\langle \langle j_E^2 \rangle \rangle 
\leq  D^2 \langle\langle j_c^2\rangle\rangle.
\eea
%
We display in Fig. \ref{fig_voltage_etur} the TURR for energy current under voltage.
As we reduce the bandwidth, we observe the suppression of TURR and the approach to TUR$_0$. 
Remarkably, this behavior can be non-monotonic with voltage.  
Before demonstrating those features with equations, 
we provide an intuition on the non-monotonic approach to TUR$_0$ in this setup.
We refer to Eq.~(\ref{eq:levitov}) and define 
\bea
g(\epsilon)&\equiv& 
\tau_0[f_L(\epsilon)(1 - f_R(\epsilon)) + f_R(\epsilon)(1 - f_L(\epsilon))] 
\nonumber\\
&-& \tau_0^2[f_L(\epsilon) - f_R(\epsilon)]^2,
\label{eq:ge}
\eea 
such that the integrand of the expression for $\langle\langle j_E^2\rangle\rangle$ 
is given, up to a constant factor, by $\epsilon^2 g(\epsilon)$ when $0\leq\epsilon\leq D$. 
As demonstrated in Figs.~\ref{noise_integrand_V}(a)-(b), when $V\lesssim D$, increasing $V$ enhances the 
combination $\epsilon^2 g(\epsilon)$ within the transmission window. However, as we continue and increase the voltage, 
$V\gg D$, the main contribution of the noise function increasingly resides at energies above $D$, thus we
observe the suppression of fluctuations with voltage.
%
The non-monotonic behavior of energy fluctuations translates to the corresponding behavior of the TURR, 
as observed in Fig. \ref{fig_voltage_etur}.
We now support this discussion quantitatively, by simplifying the TURR, Eq. (\ref{eq:28}), in various limits.

{\it High voltage.}
Based on the limit (\ref{eq:charge_zeroD}), simply dividing it by the factor of two to account for the different (asymmetric) transmission function 
adopted here, we conclude that fluctuations in the energy current decay exponentially at large voltage.
Meanwhile, in this limit the energy current approaches the constant value
\be
    \langle j_E \rangle \rightarrow \int_0^D d\epsilon\;\epsilon = \frac{D^2}{2}.
\ee
Combining these results, we see that the limit of high voltage leads the TURR to approach TUR$_0$, 
as indeed we observe in Fig.~\ref{fig_voltage_etur}. 

As for the recovery of the TUR under physical settings,
the reasoning of Sec. \ref{sec-soft} extends to energy transport, and the TURR grows linearly with $V$ at high voltage 
if e.g., soft energy cutoffs are introduced to the transmission function,  due to the existence of values of $\epsilon$ at which $\tau(\epsilon)(1-\tau(\epsilon))\neq0$. For instance, using the transmission function of Eq.~(\ref{eq:soft}), with $\tau_0=1$, $D_1=0$ and $D_2=D$, in the limit that $\gamma D\gg 1$, the TURR far from equilibrium grows with voltage as
\be 
    \beta V\langle j_c\rangle\frac{\langle\langle j_E^2\rangle\rangle}{\langle j_E\rangle^2}\xrightarrow{V\gg D, \gamma D\gg 1}\frac{4\beta V}{\gamma D}.
\ee

{\it Low voltage.}
We recall that for charge transport and using the hard cutoff transmission function, Eq. (\ref{eq:hard}), 
TUR$_2$ was immediately violated beyond equilibrium, Eq. (\ref{eq:TURRjcEq}).
In contrast, we now show that for energy current,
at small voltage, TUR$_2$ is always satisfied. 
Our analysis here is done for the boxcar function as illustrated in Fig. \ref{noise_integrand_V}.

Close to equilibrium, the Fermi functions may be approximated 
as $f_{L,R}(\epsilon) \approx f_{eq}(\epsilon) \mp \frac{V}{2}\frac{\partial f}{\partial \epsilon}\big|_{eq}$, 
where $f_{eq}(\epsilon) = (1 + e^{\beta\epsilon})^{-1}$ (we set the equilibrium Fermi energy at zero).
This allows the $V$-dependence to be pulled out of the various integrals of Eq.~(\ref{eq:levitov}), and
the TURR is found to be 
\bea
&\beta V\langle j_c\rangle&  \frac{\langle \langle j_E^2\rangle \rangle}{  \langle j_E\rangle^2}
\nonumber\\
&\xrightarrow{V\to 0}& 
2 
\frac{\int_0^D d\epsilon\;\frac{e^{\beta\epsilon}}
{(1+e^{\beta\epsilon})^2}
\int_0^D d\epsilon\;\epsilon^2\frac{e^{\beta\epsilon}}
{(1+e^{\beta\epsilon})^2}}
{\left[\int_0^D d\epsilon\;\epsilon\frac{e^{\beta\epsilon}}{(1+e^{\beta\epsilon})^2}\right]^2}.
\eea
Comparing the various integrals by means of the Cauchy-Schwarz inequality immediately reveals that the right 
hand side is greater than 2, and TUR$_2$ is satisfied at low voltage for any $D$, 
that is, even for very narrow bands.
This result is demonstrated in Fig.~\ref{fig_voltage_etur}. 

The ratio of integrals above may be evaluated in the narrow-band and wide-band limits,  showing the TURR  to converge to the temperature-independent values of $8/3$  and $\pi^2/6(\log{2})^2\approx 3.42$, respectively.
These values indeed agree with simulations shown in Fig.~\ref{fig_voltage_etur}. 

{\it Intermediate voltage.}
Between the two regimes discussed above, as the bias voltage increases, 
yet before the region in which fluctuations decay exponentially, 
it is possible to see an intermediate increase in the value of the TURR 
(as expressed by the red dashed curve in Fig.~\ref{fig_voltage_etur}). 
We attribute this non-monotonicity to the behavior of fluctuations, see Fig. \ref{noise_integrand_V}.
Here, the integrand of the energy fluctuations, which is a product of $\epsilon^2 g(\epsilon)$ with $\tau(\epsilon)$, first increases, but then decreases as we raise the voltage.
In contrast, the energy and charge currents, and consequently the entropy production, all increase monotonically with voltage.
%

\begin{figure}[t]
    \centering
    \includegraphics[width=1.0\columnwidth]{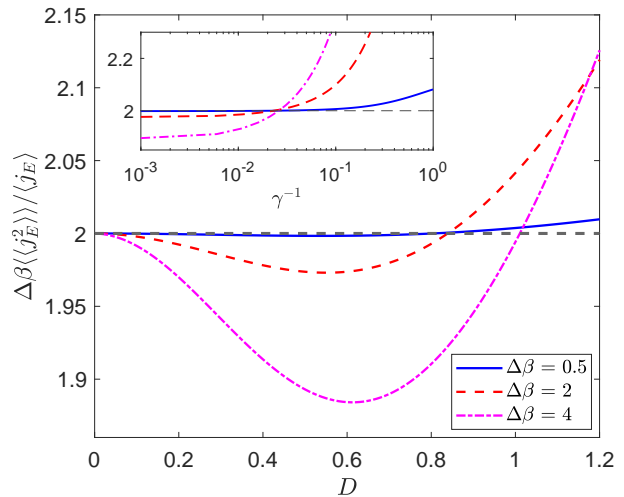}
    \caption{Energy transport TURR under a temperature difference with hard energy cutoffs, as a function of $D$. TUR$_2$ is violated at small $D$ and satisfied as $D$ grows. Inset: The TURR with soft cutoffs as a function of the broadening parameter $\gamma^{-1}$, for $D=0.6$. Violations of TUR$_2$ are observed only as the transmission function approaches a boxcar shape with hard cutoffs. $\bar{\beta}=2$, $\tau_0 = 1$.}
    \label{fig:temp_diff}
\end{figure}

\subsection{Energy transport under a temperature difference}
\label{Sec-DT}

In this Section we show that 
a temperature bias enforces a cost-precision bound on the energy current even with ideal-boxcar transmission functions, contrary to (theoretical) complete TUR violations under voltage.
We study the steady-state energy transport under a temperature difference 
described by $\Delta\beta = \beta_R - \beta_L$, with no bias voltage. 
In this case, the entropy production rate in steady state is given 
by $\langle\sigma\rangle = \Delta\beta\langle j_E\rangle$ and the TURR reduces to
$\Delta\beta\langle\langle j_E^2\rangle\rangle/\langle j_E\rangle$.

Before getting to the technical details, we provide a quantitative argument 
as to why it is impossible to approach TUR$_0$ in this scenario.
We know that under a temperature difference, the form of the difference 
$f_L(\epsilon)-f_R(\epsilon)$ departs significantly from that 
depicted in Fig.~\ref{fig1}(b) and \ref{fig1}(c), being instead a function with odd symmetry.
Significantly, we now have that $|f_L(\epsilon)-f_R(\epsilon)|\leq1/2$ everywhere, 
meaning no choice of parameters permits the 
approximation $f_L(\epsilon)-f_R(\epsilon)\approx 1$, 
and thus, $(f_L(\epsilon)-f_R(\epsilon))^2\approx f_L(\epsilon)-f_R(\epsilon)$, cannot be satisfied
throughout the region where the transmission function is nonzero. 
As a result, the $\tau^2$-contribution to the fluctuations, 
which is most significant for $\beta_L$ approaching zero and $\beta_R$ large, 
never grows large enough to effectively cancel out the $\tau$-order contribution. 
Rather than approaching zero as in the bias voltage case, the fluctuations approach a finite value at large $\Delta\beta$ 
and no analogous argument can be made for the TUR bound to approach TUR$_0$.

The TURR in our model is presented in Fig.~\ref{fig:temp_diff}. 
In accord with the discussion above, we find that TUR$_2$ may be broken for intermediate bands---with a hard cutoff---but we do not approach TUR$_0$. 
TUR$_2$ violations are also suppressed when we introduce a transmission function with soft energy cutoffs as given 
in Eq.~(\ref{eq:soft}). 
At sufficiently large broadening parameter, $\gamma^{-1}$, for a given bandwidth, these violations do not occur.
The detailed discussion of the different regions (small and large $D$, small and large $\Delta \beta$) is relegated to the Appendix.


\section{Two-affinity transport}
\label{Sec-two}

For completeness, we consider a thermoelectric junction as discussed in Ref. \cite{Landi-Boxcar1}.
We show the approach to TUR$_0$ under a small temperature bias, but again point out the fragility of the effect and the recovery of a cost-precision tradeoff in physical settings, with TUR$_2$ showing up far from equilibrium. 

\begin{figure}[t]
\vspace{7mm}
\includegraphics[width = 0.9\columnwidth, trim=30 10 30 50]{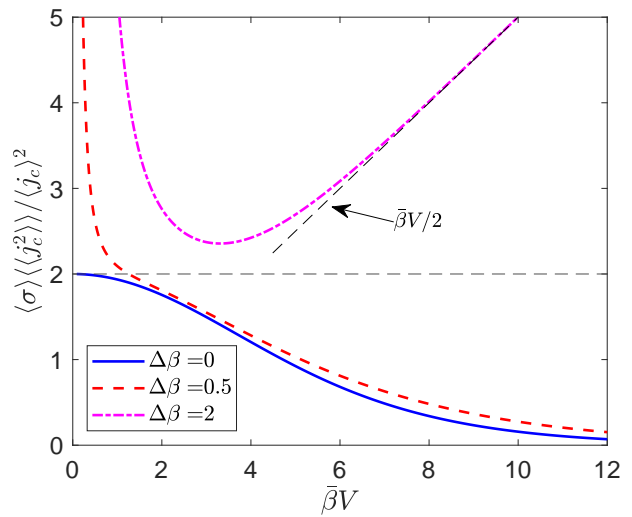}
\caption{The TUR ratio as a function of voltage for two leads at the same temperature (blue curve), as well as at slightly (red curve), and dramatically (magenta curve) different temperatures. The presence of any temperature difference leads the TUR ratio to blow up at low voltage. When the right lead is at a very high temperature, the TUR ratio takes on a simple proportionality to $V$ in the limit of high $V/D$. $D=1$, $\bar{\beta}=1$, and $\tau_0=1$.}
\label{fig:two-affinity}
\end{figure}

We consider steady-state transport in a two-terminal setup with a 
rectangular transmission function, in the presence of both a bias voltage, $V$, and a 
temperature difference, reflected by a difference in the inverse temperatures of the two leads, $\Delta\beta = \beta_L - \beta_R>0$. 
We suppose a finite band stretching from $-D$ to $D$ with $\tau_0=1$. 
Notably, there are now two distinct contributions to the entropy production, associated with each affinity,
\be 
    \langle\sigma\rangle = \bar{\beta}V\langle j_c\rangle + \Delta\beta\langle j_E\rangle,
\ee
where $\bar{\beta} = (\beta_R + \beta_L)/2$. TUR$_2$ for charge transport is thus expressed as
\be\label{eq:2-aff-tur}
    \frac{\langle\langle j_c^2\rangle\rangle}{\langle j_c\rangle^2}\big[\bar{\beta}V\langle j_c\rangle + \Delta\beta\langle j_E\rangle\big]\geq2.
\ee
When the voltage approaches zero, the charge current vanishes, but the presence of a temperature difference leads 
to a nonzero energy current and the TURR blows up, as shown in Fig. \ref{fig:two-affinity}. 
TUR$_2$ is clearly satisfied in this regime.

Moving to the opposite limit of high voltage, $V\gg D$, the behavior of the TUR ratio sensitively  depends
on the magnitude of the temperature difference between the leads. 
A small temperature difference may be seen as a perturbation on the Fermi-Dirac distributions describing the two leads 
if their temperatures were the same, for example, slightly decreasing the width of $f_L(\epsilon)$ and increasing the width of $f_R(\epsilon)$. 
However, as demonstrated in Fig.~\ref{fig:two-affinity}, under a relatively small temperature difference, 
the reasoning of Section \ref{Sec-ferm} continues to apply, and the trivial bound TUR$_0$ is approached as $V$ grows, though once nonideal conditions are introduced, TUR$_2$ is recovered at high voltage (not shown).

In the opposite scenario that the right (hot) lead approaches an 
infinite temperature, $\beta_R\rightarrow0$ and $\beta_L \approx \Delta\beta=2\bar{\beta}$, 
the TURR instead takes on a simple linear dependence with voltage in the high-$V$ limit, as we show in Fig. \ref{fig:two-affinity} (dashed-dotted line), and discuss next:

We suppose the voltage to be high enough that we may take $f_L(\epsilon)\approx1$ for $|\epsilon|\leq D$. 
However, the high temperature of the right lead necessitates that $f_R(\epsilon)\approx 1/2$. 
This leads to simplified expressions for the mean and variance of charge current,
\bea 
    \langle j_c\rangle \approx \int_{-D}^D\frac{d\epsilon}{2} = D,
    \nonumber\\
    \langle\langle j_c^2\rangle\rangle \approx \int_{-D}^D d\epsilon\bigg[\frac{1}{2} - \frac{1}{4}\bigg] = \frac{D}{2}.
\eea
The mean energy current, $\langle j_E\rangle \approx \int_{-D}^Dd\epsilon\;\epsilon/2$, 
vanishes in the high voltage limit due to odd symmetry of the integrand, 
simplifying Eq.~(\ref{eq:2-aff-tur}), leading the TUR ratio to take on the value of
\be 
    \bar{\beta}V\frac{\langle\langle j_c^2\rangle\rangle}{\langle j_c\rangle}\xrightarrow{V\gg D}\frac{\bar{\beta}V}{2}.
\ee
Thus, like in the single-affinity case with soft energy cutoffs, the TURR in this high-temperature scenario exhibits 
a linear dependence on $V$ at high voltage, guaranteeing that TUR$_2$ is ultimately satisfied in this regime. 
This is the case even if the transmission function features hard energy cutoffs, 
as is demonstrated in Fig.~\ref{fig:two-affinity} by the diagonal asymptote characterizing the dashed-dotted curve at high voltage.




\section{Summary}
\label{Sec-summ}

We have addressed the fundamental problem of whether cost-precision tradeoff relations can be completely violated in quantum coherent devices. Our answer is {\it negative for physical devices} suffering  from any degree of imperfection in their transmission profile. This imperfection translates to a nonzero noise level, and to the  thermodynamic uncertainty relation ultimately sprouting with increasing voltage.

In detail, we have studied the behavior of the TURR,  $\langle\sigma\rangle\langle\langle j_\alpha^2\rangle\rangle/\langle j_\alpha\rangle^2$, for charge and energy transport
of coherent, noninteracting electrons through a 
nanojunction under various circumstances. 
In particular, we examined TUR-violating behavior occurring in the special case that the otherwise perfect transmission function drops 
sharply to zero at energy cutoffs, and assessed the extent to which this behavior 
diminishes with  more realistic transmission functions featuring imperfect and soft energy cutoffs.
Our main observations are:

(i) The TUR for charge transport under a bias voltage is violated {\it immediately} at low voltage and approaches a trivial bound,
    $\beta V\frac{\langle\langle j_c^2\rangle\rangle}{\langle j_c\rangle}\geq0$,
at high voltage---in the idealized case of a boxcar transmission function. 
However, any deviation from this idealization restores a nontrivial cost-precision tradeoff relation. 

Furthermore, in nonideal settings and sufficiently far from equilibrium, the TURR grows linearly with voltage, ultimately recovering the TUR$_2$ bound. 
%
In the present model, the width of the tail of the transmission function plays a key role in this behavior: 
with the functions considered, for which the transmission probability drops off exponentially around the edges of the band, the TURR at high voltage is proportional to this width, set by $\gamma^{-1}$. Furthermore, the voltage range at which linear $V$-dependence begins is higher for a larger value of $\gamma$. 

(ii) The charge transport TURR exhibits a similar $V$-dependence at high voltage in the multi-affinity case, 
once again recovering TUR$_2$, given a large enough temperature difference, even with an idealized boxcar transmission function.

(iii) The behavior of the TURR for energy currents is more compound. 
Under a bias voltage, TUR$_2$ for energy transport is always satisfied at low voltage, 
in contrast to that for charge transport, which is violated.  However,
the TURR for energy transport attains behavior analogous to that for charge transport at high enough voltage. 
Under a temperature difference, limitations on the Fermi-Dirac distributions of the leads restrict the magnitude of 
TUR$_2$ violations that occur; these violations are also suppressed when a realistic transmission function with soft energy cutoffs is introduced.

Altogether, we argue that
real systems cannot avoid a (nontrivial) cost-precision tradeoff relation.
As well, while significant violations of the standard TUR$_2$ may be observed in the intermediate nonequilibrium regime when employing carefully-crafted transmission functions, in real systems TUR$_2$ is ultimately restored far from equilibrium. 
The suppression of the TURR with bandwidth, and the violation and restoration of TUR$_2$ with voltage 
could be realized experimentally in quantum transport experiments through quantum dot arrays or by simulating the junction with a cold-atom system. 

Understanding the impact of many-body interactions on the TURR in nonequilibrium steady state is left for future work.
Other related problems of interest are understanding the interrelated behavior 
of charge and energy current fluctuations, particularly when the system operates as a thermal machine far from equilibrium
\cite{Gerry21, Gerry22}, and investigating the opposite scenario of that considered in this work, that is, of nonequilibrium 
junctions displaying current noise with zero net current  \cite{Oren,Janine,Martin,Zhang}.



\begin{acknowledgments}
We acknowledge fruitful discussions with Bijay K. Agarwalla.
DS acknowledges the NSERC discovery grant and the Canada Research Chair Program. 
The research of MG was supported by an Ontario Graduate Scholarship (OGS). 
\end{acknowledgments}


\renewcommand{\theequation}{A\arabic{equation}}
\setcounter{equation}{0}  
\setcounter{section}{0} 
\section*{Appendix: TURR for energy current under temperature bias}
\label{app:1}

In this Appendix we analyze several limits for the TURR. Overall, while TUR$_2$ may be violated, we find that we cannot approach
TUR$_0$ in this setup.
For simplicity, we consider here the hard cutoff model, Eq. (\ref{eq:hard}), for the transmission probability.

\subsection{Small bandwidth limit}
First, we show that in the limit that the bandwidth $D$ approaches zero, 
the TURR approaches 2.  
We return to the symmetric-rectangle form the transmission function; 
only values of $\epsilon$ near zero contribute to the integrals for noise and current.
Since 
\bea
&&
f_L(\epsilon)(1 - f_R(\epsilon))+f_R(\epsilon)(1 - f_L(\epsilon))\approx 1/2,
\nonumber\\
&&
f_L(\epsilon) - f_R(\epsilon)\approx \Delta\beta\epsilon/4
\eea 
in this regime, 
the energy current and noise (which is dominated by the linear $\tau$-term) are given by
\bea
    \langle j_E\rangle&\approx&\frac{\Delta\beta}{4}\int_{-D}^Dd\epsilon\;\epsilon^2 = \Delta\beta\frac{D^3}{6},
    \nonumber\\
    \langle\langle j_E^2\rangle\rangle &\approx&\frac{1}{2}\int_{-D}^Dd\epsilon\;\epsilon^2 = \frac{D^3}{3}.
\eea
Cancellation of $D$ and $\Delta\beta$ occurs when putting together the TURR, 
leading to the limit, 
\bea
\Delta\beta\langle\langle j_E^2\rangle\rangle/\langle j_E\rangle\xrightarrow{D\to 0}2.
\eea
Violations of TUR$_2$ are, however, observed for finite $D$ at small $\Delta\beta$. 
Expanding expressions for the current and its fluctuations to second order in $\Delta\beta$, the TURR is approximated as
\bea
    &\Delta\beta&\frac{\langle\langle j_E^2\rangle\rangle}{\langle j_E\rangle}
    \nonumber\\
    &\approx& 2 + \bigg(\frac{\Delta\beta}{2}\bigg)^2\frac{\int_{-D}^Dd\epsilon\;\epsilon^4\frac{e^{2\bar{\beta}\epsilon}}{(1+e^{\bar{\beta}\epsilon})^4}(e^{\bar{\beta}\epsilon} - 4)}{\int_{-D}^Dd\epsilon\;\epsilon^2\frac{e^{\bar{\beta}\epsilon}}{(1+e^{\bar{\beta}\epsilon})^2}},
\label{eq:TURRE}
\eea
where $\bar{\beta} = (\beta_R - \beta_L)/2$. The integrand of the numerator in the second-order term 
is negative as long as $e^{\bar{\beta}\epsilon}-4<0$.
 Thus, for sufficiently small $D$, this integrand is negative throughout the region of 
integration and the TURR is guaranteed to take on a value less than 2, i.e., TUR$_2$ is violated.

\subsection{Large bandwidth limit}

In the limit of large $D$, however, (\ref{eq:TURRE}), which expands the TURR to second order in $\Delta \beta$, evaluates to
\bea
    \Delta\beta\frac{\langle\langle j_E^2\rangle\rangle}{\langle j_E\rangle}&\approx& 2 + \bigg(\frac{\Delta\beta}{2}\bigg)^2\frac{150\pi^2 - 7\pi^4 + 810\zeta(3)}{15\pi^2\bar{\beta}^2}
    \nonumber\\
    &\approx& 2 + 2.99\bigg(\frac{\Delta\beta}{\bar{\beta}}\bigg)^2.
\eea
The second-order term is always positive, guaranteeing that at small thermal bias $\Delta\beta$ 
and large $D$, TUR$_2$ is always satisfied. 
This behavior is  confirmed by simulations as $\Delta\beta$ grows, as shown in Fig.~\ref{fig:temp_diff}.

\subsection{Large temperature gradient}
Finally, we  consider the case where the temperature of one lead is very high, taking $\beta_L\rightarrow0$, and 
study the behavior of the energy current and its noise as $\beta_R$ grows. 
In this case, $f_L(\epsilon)\approx1/2$ everywhere and $f_L(\epsilon)-f_R(\epsilon)\approx\tanh(\beta_R\epsilon/2)/2$. 
This quantity is linear in $\epsilon$ for small $\epsilon$, and approximately equal to $1/2$ as 
long as $\beta_R\epsilon\gg1$. Thus, for $\beta_RD\gg1$, as will often be the case for large $\beta_R$, $f_L(\epsilon)-f_R(\epsilon)$ may be treated to a good approximation as a constant ($1/2$) throughout the region of integration, leaving
\be
    \langle j_E\rangle \approx \int_{-D}^Dd\epsilon\;\frac{\epsilon}{2}=\frac{D^2}{2}.
\ee
Similarly, in this limit, $f_L(\epsilon) + f_R(\epsilon) - 2f_L(\epsilon)f_R(\epsilon) \approx 1/2$, 
therefore the integrand of the expression for $\langle\langle j_E^2\rangle\rangle$ approaches a quadratic 
function, independent of $\beta_R$ (see Fig.~\ref{fig:temp_diff}), and
\bea
    \langle\langle j_E^2\rangle\rangle &\approx& \int_{-D}^D d\epsilon\; \epsilon^2\left[\frac{1}{2} 
- \frac{1}{4}\tanh^2\left(\frac{\beta_R\epsilon}{2}\right)\right]
    \nonumber\\
    &\approx& \int_{-D}^Dd\epsilon\; \frac{\epsilon^2}{4} = \frac{D^3}{6}.
\eea
In the limit of $\beta_L\rightarrow0$ and $\beta_RD\gg1$, the TURR is linear in D,
\be
    \Delta\beta\frac{\langle\langle j_E^2\rangle\rangle}{\langle j_E\rangle}\xrightarrow[]{\beta_L\rightarrow0,~\beta_RD\gg1}\beta_R\frac{D}{3},
\ee
and TUR$_2$ is satisfied, since $\beta_RD\gg1$.

Conversely, supposing one lead is at a very low temperature, 
for instance taking $\beta_R\rightarrow\infty$ and leaving $\beta_L$ finite,
 we may take $f_R(\epsilon)=0$ for $\epsilon>0$ and $1 - f_R(\epsilon) = 0$ for $\epsilon<0$. Therefore,
throughout the region of integration the energy current and its variance simplify to
\bea
    \langle j_E\rangle &=& \int_{-D}^Dd\epsilon\;\epsilon f_L(\epsilon),
    \nonumber\\
    \langle\langle j_E^2\rangle\rangle &=& \int_{-D}^Dd\epsilon\;\epsilon^2f_L(\epsilon)(1-f_L(\epsilon)).
\eea
These integrals are both positive and finite, even as $D\rightarrow\infty$, in which case we have $\langle\langle j_E^2\rangle\rangle/\langle j_E\rangle\rightarrow 2/\beta_L$. Thus, the TURR 
blows up in this limit, since $\Delta\beta = \beta_R - \beta_L \rightarrow\infty$.



\begin{thebibliography}{4}

\bibitem{Seifert-Rev20}
U. Seifert, From Stochastic Thermodynamics to Thermodynamic Inference, Annu. Rev. Condens. Matter Phys. {\bf 10}, 171 (2019).

\bibitem{Horowitz20}
J. M. Horowitz and T. R. Gingrich,
Thermodynamic uncertainty relations constrain non-equilibrium fluctuations,
Nature Physics {\bf 16}, 15 (2020).

\bibitem{Barato:2015:UncRel}
A. C. Barato and U. Seifert,
Thermodynamic uncertainty relation for biomolecular processes,
Phys. Rev. Lett. {\bf 114}, 158101 (2015).

\bibitem{Gingrich:2016:TUP}
T. R. Gingrich, J. M.  Horowitz, N. Perunov, and J. L. England,
Dissipation bounds all steady-state current fluctuations,
Phys. Rev. Lett. {\bf 116}, 120601 (2016).

\bibitem{Horowitz:2017:TUR}
J. M. Horowitz and T. R. Gingrich,
Proof of the finite-time thermodynamic uncertainty relation for steady-state currents,
Phys. Rev. E {\bf 96}, 020103(R) (2017).

\bibitem{Dechant:2018:TUR}
A. Dechant,
Multidimensional thermodynamic uncertainty relations,
J. Phys. A: Math. Theor. {\bf 52}, 035001 (2019).

\bibitem{Pietzonka:2017:FiniteTUR}
P. Pietzonka, F. Ritort, and U. Seifert,
Finite-time generalization of the thermodynamic uncertainty relation,
Phys. Rev. E {\bf 96}, 012101 (2017).

\bibitem{Pigolotti:TURF}
S. Pigolotti, I. Neri, E. Roldán, and F. J\"ulicher,
Generic Properties of Stochastic Entropy Production,
Phys. Rev. Lett. {\bf 119}, 140604 (2017).

\bibitem{Gingrich:2017}
T. R. Gingrich, G. M. Rotskoff and J. M Horowitz,
Inferring dissipation from current fluctuations,
J. Phys. A: Math. Theor. {\bf 50} 184004 (2017). 

\bibitem{Hasegawa1}
Y. Hasegawa and T. V. Vu,
Uncertainty relations in stochastic processes: An information inequality approach,
Phys. Rev. E {\bf 99}, 062126  (2019). 

\bibitem{TUR-gupta}
D. Gupta and A. Maritan,
Thermodynamic uncertainty relations via second law of thermodynamics,
Eur. Phys. J. B {\bf 93}, 28 (2020). 

\bibitem{Hyeon:2017:TUR}
C. Hyeon and W. Hwang,
Physical insight into the thermodynamic uncertainty relation using Brownian motion in tilted periodic potentials,
Phys. Rev. E {\bf 96}, 012156 (2017).


\bibitem{Koyuk:2018:PeriodicTUR}
T. Koyuk, U. Seifert, and P. Pietzonka,
A generalization of the thermodynamic uncertainty relation to periodically driven systems,
J. Phys. A: Math. Theor. {\bf 52}, 02LT02 (2018).

\bibitem{Gabri}
A. C. Barato, R. Chetrite, A. Faggionato, and D. Gabrielli,
Bounds on current fluctuations in periodically driven systems,
New J. Phys. {\bf 20}, 103023 (2018).

\bibitem{Esposito20}
G. Falasco, M. Esposito, and J. C. Delvenne,
Unifying thermodynamic uncertainty relations,
New Journal of Physics {\bf 22}, 053046 (2020).

\bibitem{Potanina}
E. Potanina, C. Flindt, M. Moskalets, and K. Brandner,
Thermodynamic bounds on coherent transport in periodically driven conductors,
Phys. Rev. X {\bf 11}, 021013 (2021).

\bibitem{GarrahanLR-broken}
K. Macieszczak, K. Brandner, and J. P. Garrahan,
Unified Thermodynamic Uncertainty Relations in Linear Response, Phys. Rev. Lett. {\bf 121}, 130601 (2018).


\bibitem{Udo:TURB}
H.-M. Chun, L. P. Fischer, and U. Seifert,
Effect of a magnetic field on the thermodynamic uncertainty relation,
Phys. Rev. E {\bf 99}, 042128 (2019).

\bibitem{Saito}
K. Brandner, T. Hanazato, and K. Saito,
Thermodynamic bounds on precision in ballistic multiterminal transport,
Phys. Rev. Lett. {\bf 120}, 090601 (2018).


\bibitem{Hyst}
K. Proesmans and J. M. Horowitz,
Hysteretic thermodynamic uncertainty relation for systems with broken time-reversal symmetry,
J. Stat. Mech. 054005 (2019).



\bibitem{Landi-PRL}
A. M. Timpanaro, G. Guarnieri, J. Goold, and G. T. Landi,
Thermodynamic uncertainty relations from exchange fluctuation theorems,
Phys. Rev. Lett. {\bf 123}, 090604 (2019).

\bibitem{VanTUR}
Y. Hasegawa and T. Van Vu,
Fluctuation theorem uncertainty relation,
Phys. Rev. Lett. {\bf 123}, 110602 (2019). 

\bibitem{fluctRev}
M. Esposito, U. Harbola, and S. Mukamel,
Nonequilibrium fluctuations, fluctuation theorems, and counting statistics in quantum systems,
Rev. Mod. Phys. {\bf 81}, 1665 (2009).




\bibitem{Ptasz}
K. Ptaszynski, Coherence-enhanced constancy of a quantum thermoelectric generator, 
Phys. Rev. B {\bf 98}, 085425 (2018).

\bibitem{BijayTUR}
B. K. Agarwalla and D.  Segal,
Assessing the validity of the thermodynamic uncertainty relation in quantum systems,
Phys. Rev. B  {\bf 98}, 155438 (2018).

\bibitem{LandiQ}
G. Guarnieri, G. T. Landi, S. R. Clark, J. Goold,
Thermodynamics of precision in quantum non equilibrium steady states,
Phys. Rev. Research {\bf 1}, 033021 (2019).

\bibitem{Samuelsson}
S. Kheradsoud, N. Dashti, M. Misiorny, P. Potts, J. Splettstoesser, and P. Samuelsson,
Power, Efficiency and Fluctuations in a Quantum Point Contact as Steady-State Thermoelectric Heat Engine,
Entropy {\bf 21}, 777 (2019).

\bibitem{BijayH}
S. Saryal, H. Friedman, D. Segal, and B. K. Agarwalla,
Thermodynamic uncertainty relation in thermal transport,
Phys. Rev. E {\bf 100}, 042101 (2019).

\bibitem{Junjie}
J. Liu and D. Segal,
Thermodynamic uncertainty relation in quantum thermoelectric junctions,
Phys. Rev. E {\bf 99}, 062141 (2019).


\bibitem{Hava}
H. M. Friedman, B. K. Agarwalla, O. Shein-Lumbroso, O. Tal, and D. Segal,
Thermodynamic uncertainty relation in atomic-scale quantum conductors,
Phys. Rev. B {\bf 101}, 195423 (2020). 

\bibitem{Liu:coh}
J. Liu and D. Segal,
Coherences and the thermodynamic uncertainty relation: Insights from quantum absorption refrigerators,
Phys. Rev. E {\bf 103}, 032138 (2021).

\bibitem{Goold21}
A. Rignon-Bret, G. Guarnieri, J. Goold, and M. T. Mitchison,
Thermodynamics of precision in quantum nanomachines,
Phys. Rev. E {\bf 103}, 012133 (2021).

\bibitem{Arash}
A. Arash Sand Kalaee, A. Wacker, and P. P. Potts,
Violating the thermodynamic uncertainty relation in the three-level maser,
Phys. Rev. E {\bf 104}, L012103 (2021).

\bibitem{Gernot21}
T. Ehrlich and G. Schaller,
Broadband frequency filters with quantum dot chains,
Phys. Rev. B {\bf 104}, 045424 (2021),


\bibitem{Landi-Boxcar1}
A. M. Timpanaro, G. Guarnieri, and G. T. Landi,
The most precise quantum thermoelectric,
arXiv:2108.05325.
 
\bibitem{Landi-Boxcar2}
A. M. Timpanaro, G. Guarnieri, and G. T. Landi,
Hyperaccurate thermoelectric currents, 
 	arXiv:2108.05325.


\bibitem{Gerry21}
S. Saryal, M. Gerry, I. Khait, D. Segal, and B. K. Agarwalla,
Universal Bounds on Fluctuations in Continuous Thermal Machines,
Phys. Rev. Lett. {\bf 127}, 190603 (2021).

\bibitem{Gerry22}
M. Gerry, N. Kalantar, and D. Segal,
Bounds on fluctuations for ensembles of quantum thermal machines,
arXiv:2109.03526.


\bibitem{Carollo19}
F. Carollo, R. L. Jack, and J. P. Garrahan, 
Unraveling the Large Deviation Statistics of Markovian Open Quantum Systems, 
Phys. Rev. Lett. {\bf 122}, 130605 (2019).


\bibitem{Hasegawa21a}
Y. Hasegawa, 
Thermodynamic Uncertainty Relation for General Open Quantum Systems, 
Phys. Rev. Lett. {\bf 126}, 010602 (2021).

\bibitem{Hasegawa21b}
Y. Hasegawa,
Irreversibility, Loschmidt Echo, and Thermodynamic Uncertainty Relation,
Phys. Rev. Lett. {\bf 127}, 240602 (2021).

\bibitem{Saito-OQS}
T. Van Vu and K. Saito,
Thermodynamics of Precision in Markovian Open Quantum Dynamics, arXiv:2111.04599.

\bibitem{Levitov} L. S. Levitov and G. B. Lesovik, Charge distribution in quantum shot noise, Pis’ma Zh. Eksp. Teor. Fiz. {\bf 58}, 225 (1993) [JETP Lett. {\bf 58}, 230 (1993)].

\bibitem{fcs-charge1}
K. Sch\"onhammer, Full counting statistics for noninteracting fermions: Exact results and the Levitov-Lesovik formula, Phys. Rev. B {\bf 75}, 205329 (2007).

\bibitem{fcs-charge2}
B. K. Agarwalla, B. Li, and J.-S. Wang, Full-counting statistics of heat transport in harmonic junctions: Transient, steady states,
and fluctuation theorems, Phys. Rev. E {\bf 85}, 051142 (2012).


\bibitem{Nitzan}
A. Nitzan,
{\it Chemical Dynamics in Condensed Phases: Relaxation, Transfer and Reactions in Condensed Molecular Systems},
(Oxford University Press, UK 2006).

\bibitem{diventra}
M. Di Ventra,
{\it Electrical Transport in Nanoscale Systems}, (Cambridge University Press, Cambridge, U.K., 2008).

\bibitem{Oren}
O. Shein Lumbroso, L. Simine, A. Nitzan, D. Segal, and O. Tal,
Electronic noise due to temperature difference in atomic-scale junctions, Nature  {\bf 562}, 240 (2018). 


\bibitem{Janine}
J. Eriksson, M. Acciai, L. Tesser, and J. Splettstoesser,
General Bounds on Electronic Shot Noise in the Absence of Currents,
Phys. Rev. Lett. {\bf 127}, 136801 (2021).

\bibitem{Martin}
A. Popoff, J. Rech, T. Jonckheere, L. Raymond,
B. Gremaud, S. Malherbe, and T. Martin,
Scattering theory of non-equilibrium noise and delta-T current
fluctuations through a quantum dot, arXiv:2106.05679.

\bibitem{Zhang}
G. Zhang, I. V. Gornyi, and C. Spånslätt,
Does delta-T noise probe quantum statistics?
arXiv:2201.13174.

\end{thebibliography}
\end{document}